\documentclass[letterpaper]{jpconf}
\usepackage{graphicx}
\begin{document}
\title{Fluctuation Results from PHENIX}

\author{J.T. Mitchell and the PHENIX Collaboration}

\address{Brookhaven National Laboratory, P.O. Box 5000, Building 510C, Upton, NY 11973-5000}

\ead{mitchell@bnl.gov}

\begin{abstract}
The PHENIX Experiment at the Relativistic Heavy Ion Collider has made measurements of event-by-event fluctuations in the net charge, the mean transverse momentum, and the charged particle multiplicity as a function of collision energy, centrality, and transverse momentum in heavy ion collisions. The results of these measurements will be reviewed and discussed.
\end{abstract}.

\section{Introduction}

The PHENIX experiment at the Relativistic Heavy Ion Collider has made measurements of several event-by-event fluctuation observables. These measurements include the following: fluctuations in the net charge, which are sensitive to the charge distribution within the collision volume; fluctuations in the mean transverse momentum, which are sensitive to critical fluctuations in the temperature of the system and may be large in the presence of a phase transition from hadronic matter to a Quark-Gluon Plasma (QGP); and fluctuations in charged particle multiplicity, which appear to exhibit non-monotonic behavior as a function of centrality at SPS energies.

However, each of these fluctuation observables are also sensitive to known processes such as charge conservation, resonance decays, elliptic flow, and hard processes. Therefore the contribution of these known processes must be thoroughly investigated before reliably interpreting the results of the measurement.

Details about the PHENIX experimental configuration can be found elsewhere \cite{phenixNIM}. All of the measurements described here utilized the PHENIX central arm detectors. The PHENIX acceptance with a maximum acceptance of $|\eta|<0.35$ in pseudorapidity and $180^{o}$ in azimuthal angle is considered small for event-by-event measurements. However, the event-by-event multiplicities are high enough in RHIC heavy ion collisions that PHENIX has a competitive sensitivity for the detection of fluctuation signals. For example, a detailed examination of the PHENIX sensitivity to temperature fluctuations derived from the measurement of mean $p_{T}$ fluctuations is described in \cite{ppg005}.

\section{Net Charge Fluctuations}

A promising signature for the presence of a QGP is the observation of event-by-event fluctuations in the net charge.  It has been hypothesized that fluctuations in the net charge would be significantly reduced in a QGP scenario, where the fractional electric charges of the quarks are more evenly spread throughout the QGP volume than the unit electric charges of the hadrons in a hadronic gas volume \cite{Jeo00}. PHENIX has measured net charge fluctuations in Au+Au collisions at $\sqrt{s_{NN}}$ = 130 and 200 GeV \cite{phxQfluc130,phxQfluc200}. The measurements have been made as a function of centrality, pseudorapidity range, and azimuthal angle range. The fluctuations are quoted using the variable $v(Q)=V(Q)/<N_{ch}>$ where $Q=N_{+}-N_{-}$ is the net charge, $N_{ch}=N_{+}+N_{-}$ is the total number of charged particles, and $V(Q)$ represents the variance of the net charge. The expectations for the value of $v(Q)$ are 1.0 for random particle emission, 0.75 for a resonance gas, 0.83 for a quark coalescence scenario, and 0.25 for a QGP in small acceptance ranges \cite{Bia02} . 

\begin{figure}[h]
\begin{minipage}{18pc}
\includegraphics[width=18pc]{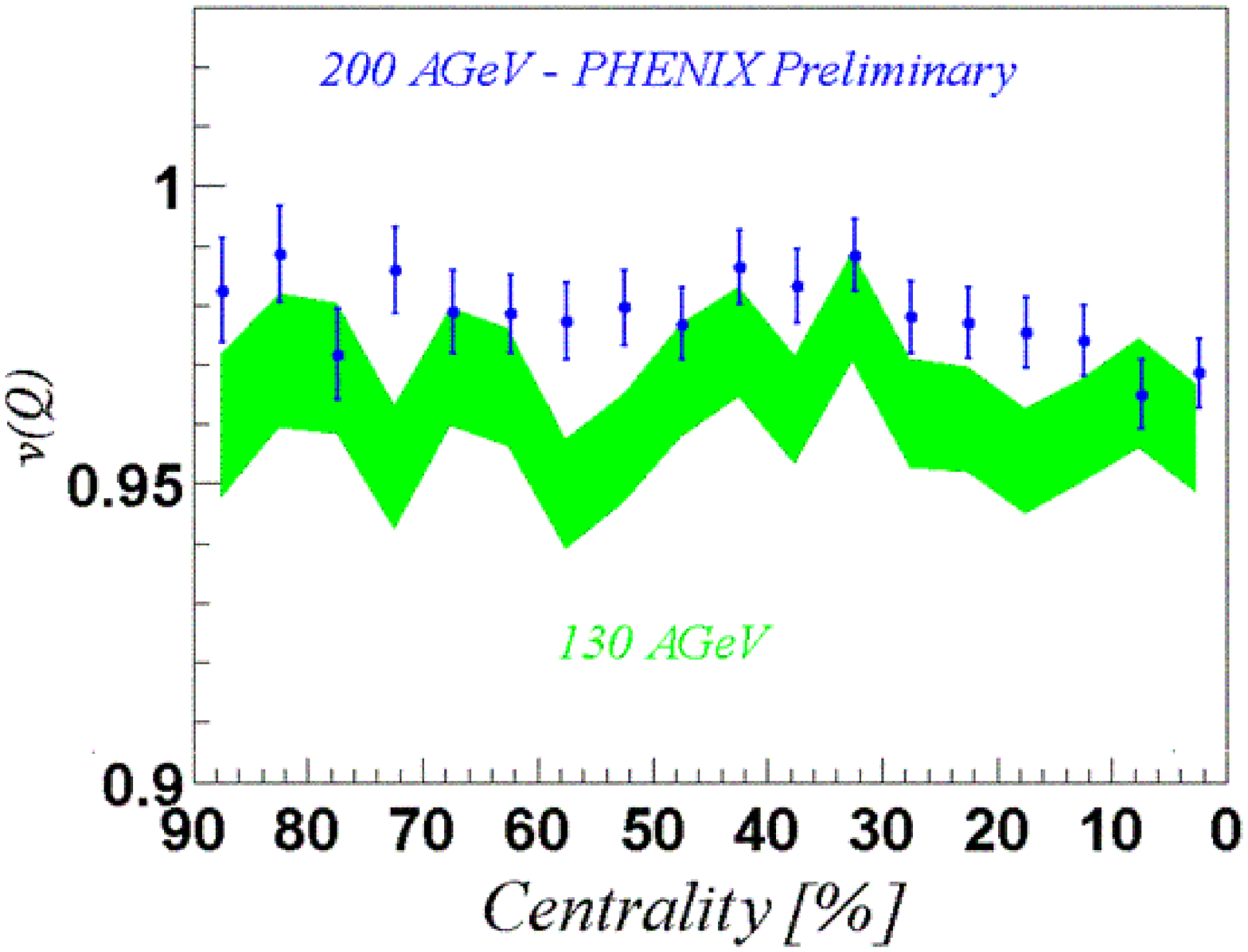}
\caption{\label{fig:qflucVsCent}The charge fluctuation variable $v(Q)$ as a function of centrality for $\sqrt{s_{NN}}$ = 130 GeV (band) and $\sqrt{s_{NN}}$ = 200 GeV (points) Au+Au collisions. Here, peripheral to central collisions proceed from left to right. The width of the band represents the statistical error from the 130 GeV measurement. The acceptance range is $|\eta|<0.35$, $p_{T}>200$ MeV/c, and $\Delta \phi = \pi/2$ in azimuth.}
\end{minipage}\hspace{2pc}%
\begin{minipage}{18pc}
\includegraphics[width=18pc]{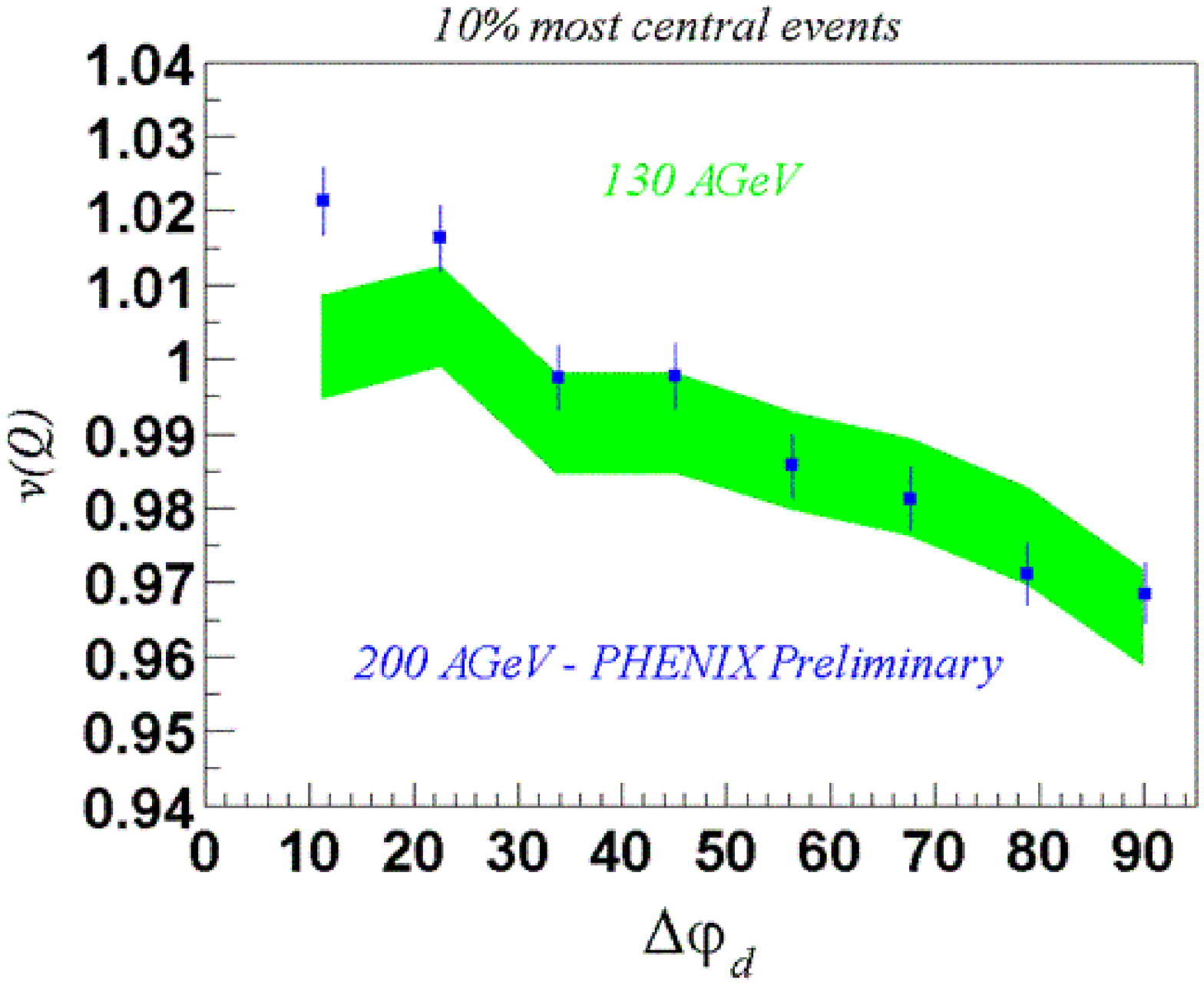}
\caption{\label{fig:qflucVsPhi}$v(Q)$ as a function of azimuthal angle range for $\sqrt{s_{NN}}$ = 130 GeV (band) and $\sqrt{s_{NN}}$ = 200 GeV (points) Au+Au 0-10\% central collisions. The width of the band represents the statistical error from the 130 GeV measurement.}
\end{minipage} 
\end{figure}

Shown in Fig. \ref{fig:qflucVsCent} are the PHENIX measurements as a function of centrality.  The small difference in the two datasets is due to the effect of tighter track cuts that effectively reduce the acceptance in the $\sqrt{s_{NN}}$ = 200 GeV sample. For the 10\% most central collisions with $|\eta|<0.35$, $p_{T}>200$ MeV/c, and $\Delta \phi = \pi/2$, $v(Q)=0.965 \pm 0.007(stat.) \pm 0.019(syst.)$ for $\sqrt{s_{NN}}$ = 130 GeV and $v(Q)=0.969 \pm 0.006(stat.) \pm 0.020(syst.)$ (PHENIX Preliminary) for $\sqrt{s_{NN}}$ = 200 GeV Au+Au collisions.

The value of $v(Q)$ must be adjusted for charge conservation by a factor of $1-p$, where $p$ is the fraction of the charged particles produced within the acceptance range. $v(Q)$ is also expected to be reduced by the decay of neutral resonances. The reduction due to resonance decays should decrease with increasing geometrical acceptance since more particle pairs from larger opening angles will be included in the measurement. Fig. \ref{fig:qflucVsPhi} shows the measurement of $v(Q)$ as a function of azimuthal angle range for the 10\% most central collisions. The expected decrease in $v(Q)$ as the azimuthal range is increased is observed. The decrease is larger than the expectation from charge conservation alone by more than 3 standard deviations. However, the decrease is consistent with results from RQMD simulations \cite{phxQfluc130} in the PHENIX acceptance, thus emphasizing the importance of the resonance decay contribution. As a result, PHENIX measurements of charge fluctuations are consistent with the expectations of statistically independent particle emission.

\section{Mean Transverse Momentum Fluctuations}

\begin{figure}[h]
\begin{minipage}{18pc}
\includegraphics[width=18pc]{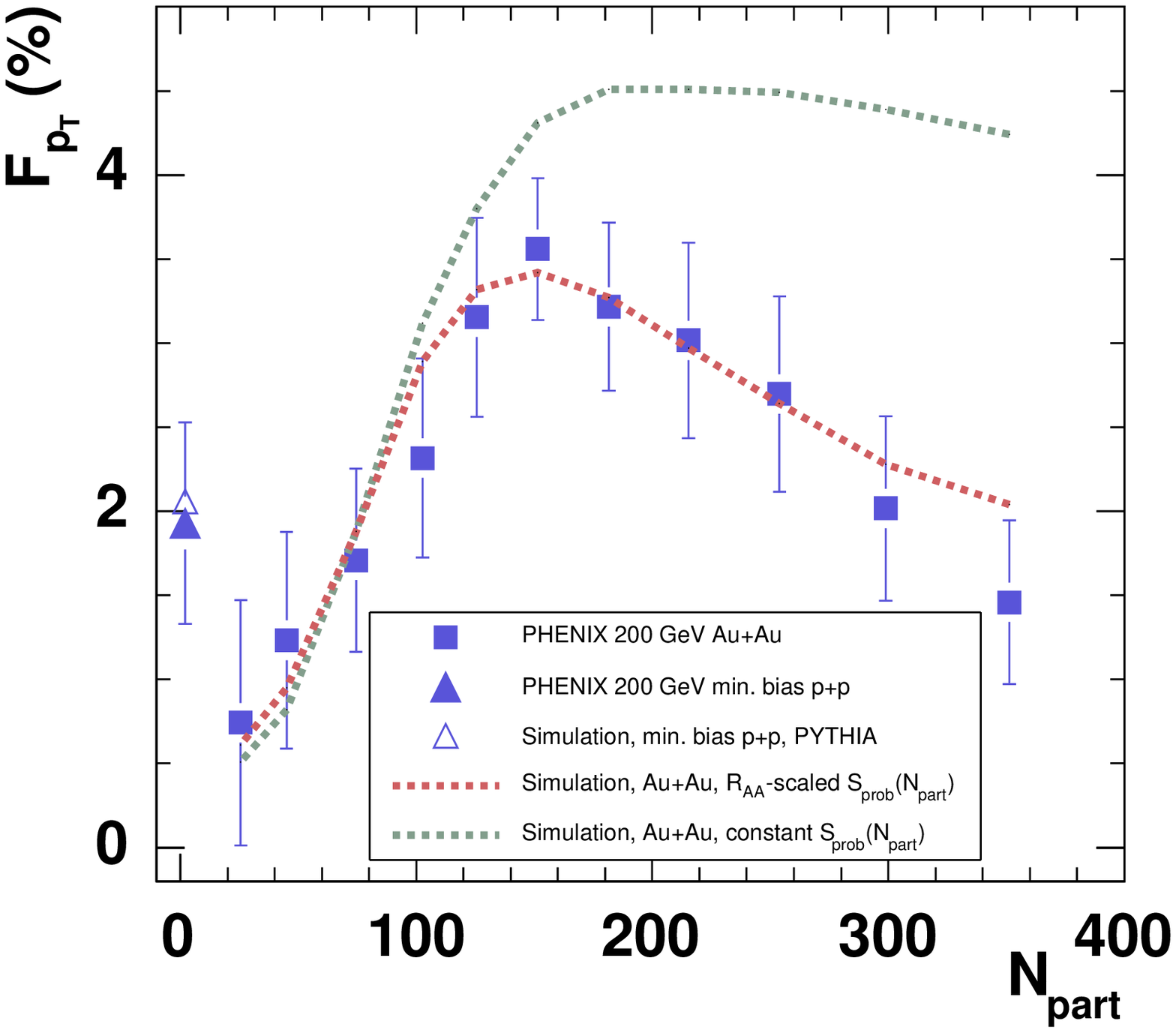}
\caption{\label{fig:ptFlucVsCentMM}Mean $p_T$ fluctuations as a function of centrality for $\sqrt{s_{NN}}$ = 200 GeV Au+Au collisions. The curves are the results of a model that estimates the relative contribution due to jet production as a function of centrality.}
\end{minipage}\hspace{2pc}%
\begin{minipage}{18pc}
\includegraphics[width=18pc]{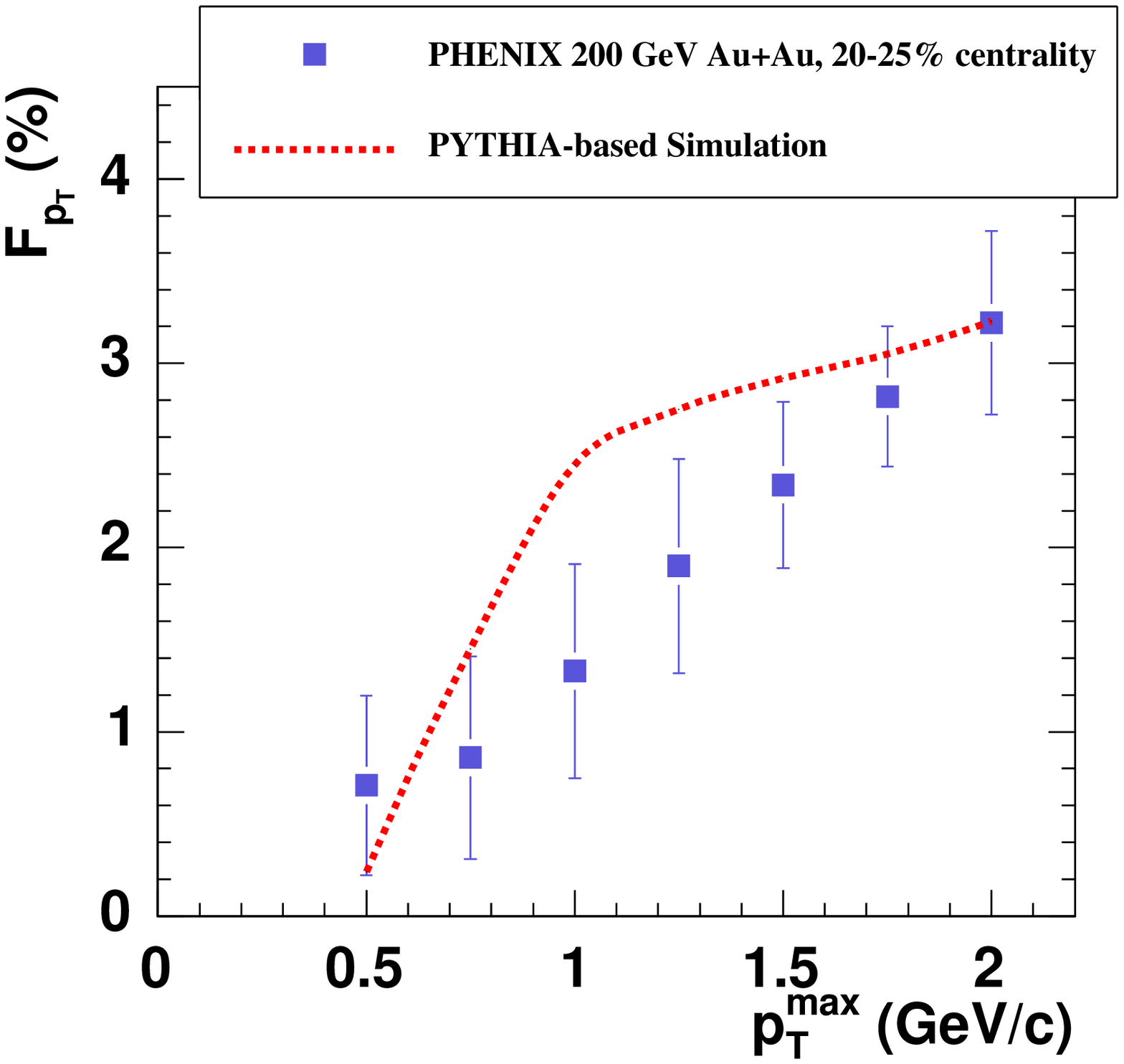}
\caption{\label{fig:ptFlucVsPtMM}Mean $p_T$ fluctuations as a function of $p_{T}$ range ($200~MeV/c<p_{T}<p_{T,max}$) for $\sqrt{s_{NN}}$ = 200 GeV Au+Au collisions. The curve is the result of a PYTHIA-based model that estimates the relative contribution due to jet production as a function of centrality.}
\end{minipage} 
\end{figure}

Event-by-event fluctuations in the mean transverse momentum, $M_{p_{T}}$, may be sensitive to temperature fluctuations near the QCD critical point \cite{Ste99}.  PHENIX has measured the magnitude of these fluctuations in Au+Au collisions at $\sqrt{s_{NN}}$ = 130 GeV and 200 GeV as a function of centrality and $p_{T}$ \cite{ppg005,ppg027}. PHENIX measurements are quoted in the variable $F_{p_T} = (\omega_{data}-\omega_{baseline})/\omega_{baseline}$ in percent, where $\omega = \sigma_{M_{p_T}}/<N>$. This observable can be related to other fluctuation variables currently in use with some reasonable assumptions \cite{Gav04}.  Here, the baseline, random, distribution is constructed using mixed events. It is important that the number of tracks, $N_{tracks}$, for the mixed events are sampled directly from the data distribution since the standard deviation of the distribution of the event-by-event mean $p_T$ depends on $N_{tracks}$. Also, no two tracks from the same data event are allowed to populate the same mixed event. $F_{p_T}$ as a function of centrality is shown in Fig. \ref{fig:ptFlucVsCentMM}. A significant non-random fluctuation is observed for all centralities.

\begin{figure}[h]
\begin{minipage}{18pc}
\includegraphics[width=18pc]{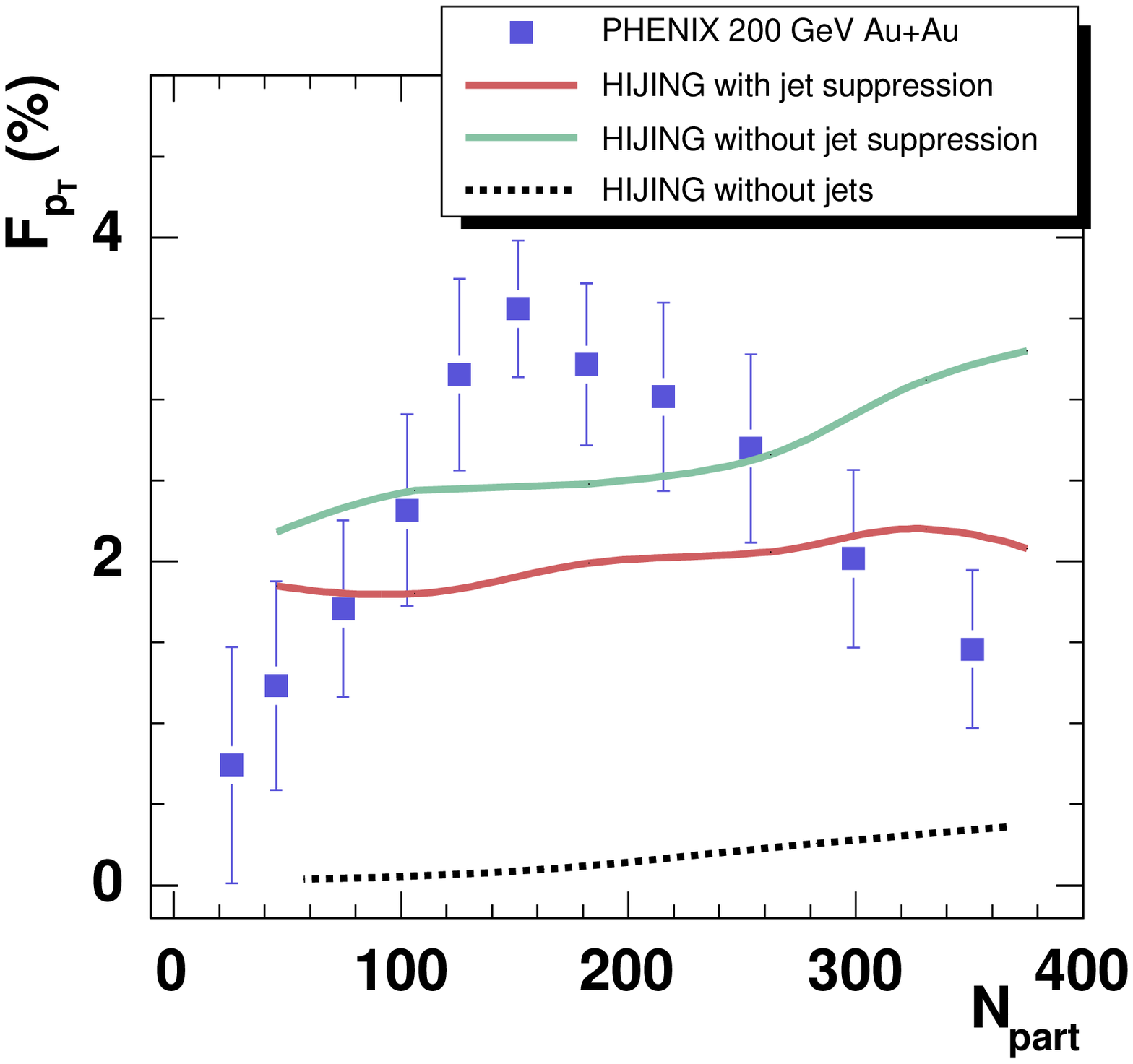}
\caption{\label{fig:ptFlucVsCentHij}Mean $p_T$ fluctuations as a function of centrality for $\sqrt{s_{NN}}$ = 200 GeV Au+Au collisions. The curves are the results of HIJING 1.35 (default parameters) with jet suppression, without jet suppression, and without jets into the PHENIX acceptance.}
\end{minipage}\hspace{2pc}%
\begin{minipage}{18pc}
\includegraphics[width=18pc]{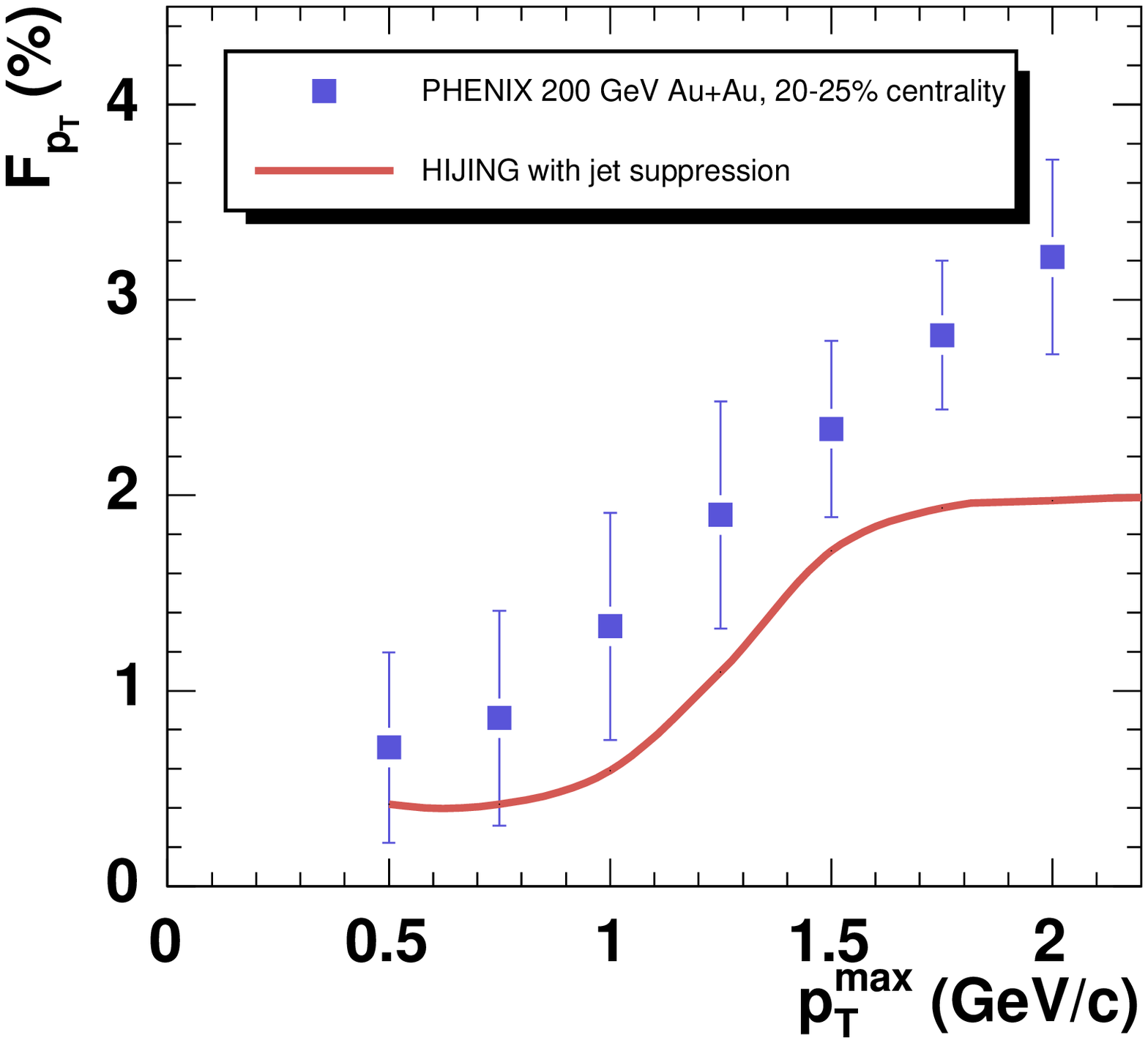}
\caption{\label{fig:ptFlucVsPtHij}Mean $p_T$ fluctuations as a function of $p_{T}$ range ($200~MeV/c<p_{T}<p_{T,max}$) for $\sqrt{s_{NN}}$ = 200 GeV Au+Au collisions. The curves are the results of HIJING 1.35 with jet suppression into the PHENIX acceptance.}
\end{minipage} 
\end{figure}

Since a positive signal is observed, contributions to the signal from known sources must first be understood.  Contributions from HBT, resonance decays, and elliptic flow have been studied using simulations and are found to be very small or negligible. Another possible source of the signal is from hard scattering processes, which are expected to produce event-by-event correlations in $p_T$, especially at higher $p_T$.  As shown in Fig. \ref{fig:ptFlucVsPtMM}, the majority of the contribution to the fluctuations occur at high $p_T$.  There is a 243\% increase in the signal when the maximum of the $p_T$ range over which the average $p_T$ is calculated increases from 1.0 to 2.0 GeV/c.  This very large increase cannot be attributed to statistical fluctuations, since the number of particles increases by less than 15\% over that range.  In order to investigate the contributions due to jet production, a data-driven two-component model consisting of 1) a simulation of soft processes by reproducing the inclusive $p_T$ and $N_{tracks}$ distributions as a function of centrality, and 2) a simulation of hard scattering processes by embedding PYTHIA jet events at a given rate per produced particle from step 1 (while conserving the $N_{tracks}$ distribution).  This rate is the only free parameter in the simulation. After adjusting the rate so that the fluctuations match the data in a reference centrality bin, the simulation is run as a function of centrality within the PHENIX detector acceptance with a) the PYTHIA event embedding rate kept constant and b) with the PYTHIA event embedding rate scaled by the measurement of the nuclear modification factor, $R_{AA}$. The results of the simulation superimposed with the PHENIX data are shown in Fig. \ref{fig:ptFlucVsCentMM} and \ref{fig:ptFlucVsPtMM}. The simulation with the $R_{AA}$-scaled rate agrees with the centrality-dependent (and also the $p_T$-dependent) trends remarkably well.  Within this simulation, the decrease of fluctuations in the most peripheral collisions is attributed to the fact that the signal begins to compete with the magnitude of the statistical fluctuations, while most of the decrease of fluctuations in the most central collisions may be attributed to the onset of jet suppression. Using the rate determined from the PHENIX data, the simulation also reproduces the magnitude of fluctuations observed by STAR \cite{starPtFluc} in 130 GeV Au+Au collisions \cite{jtmQM2004}. Comparisons to the HIJING model are shown in Figs. \ref{fig:ptFlucVsCentHij} and \ref{fig:ptFlucVsPtHij}. Although HIJING qualitatively reproduces the $p_T$-dependent trend, it does not reproduce the centrality dependence. The reason for this is not yet understood.

One way to directly compare the fluctuation measurements from all of the experiments is to estimate the magnitude of any residual event-by-event temperature fluctuations using the prescription outlined in \cite{Kor01}.  The most central collision fluctuation values correspond to maximum temperature fluctuations of 1.8\% in PHENIX, 1.7\% in STAR, 1.3\% in CERES, and 0.7\% in NA49.  Disagreement in the CERES and NA49 measurements may be due to differences in the pseudorapidity region over which the measurement was made. In all cases, the residual temperature fluctuations are small and do not significantly increase between SPS and RHIC energies.

\section{Charged Particle Multiplicity Fluctuations}

Interest in the topic of charged particle multiplicity fluctuations has been recently revived by the observation of non-monotonic behavior in the scaled variance as a function of system size at SPS energies \cite{na49MF}.  The scaled variance is defined as $var(N)/<N>$, where $var(N)$ represents the variance of the multiplicity distribution in a given centrality bin, and $<N>$ is the mean of the distribution. The scaled variance of a Poisson distribution is 1.0, independent of N. PHENIX has studied the behavior of charged particle multiplicity fluctuations as a function of centrality in $\sqrt{s_{NN}}$ = 62 GeV and 200 GeV Au+Au collisions in order to investigate if this behavior persists at RHIC energies.

\begin{figure}[h]
\begin{minipage}{18pc}
\includegraphics[width=18pc]{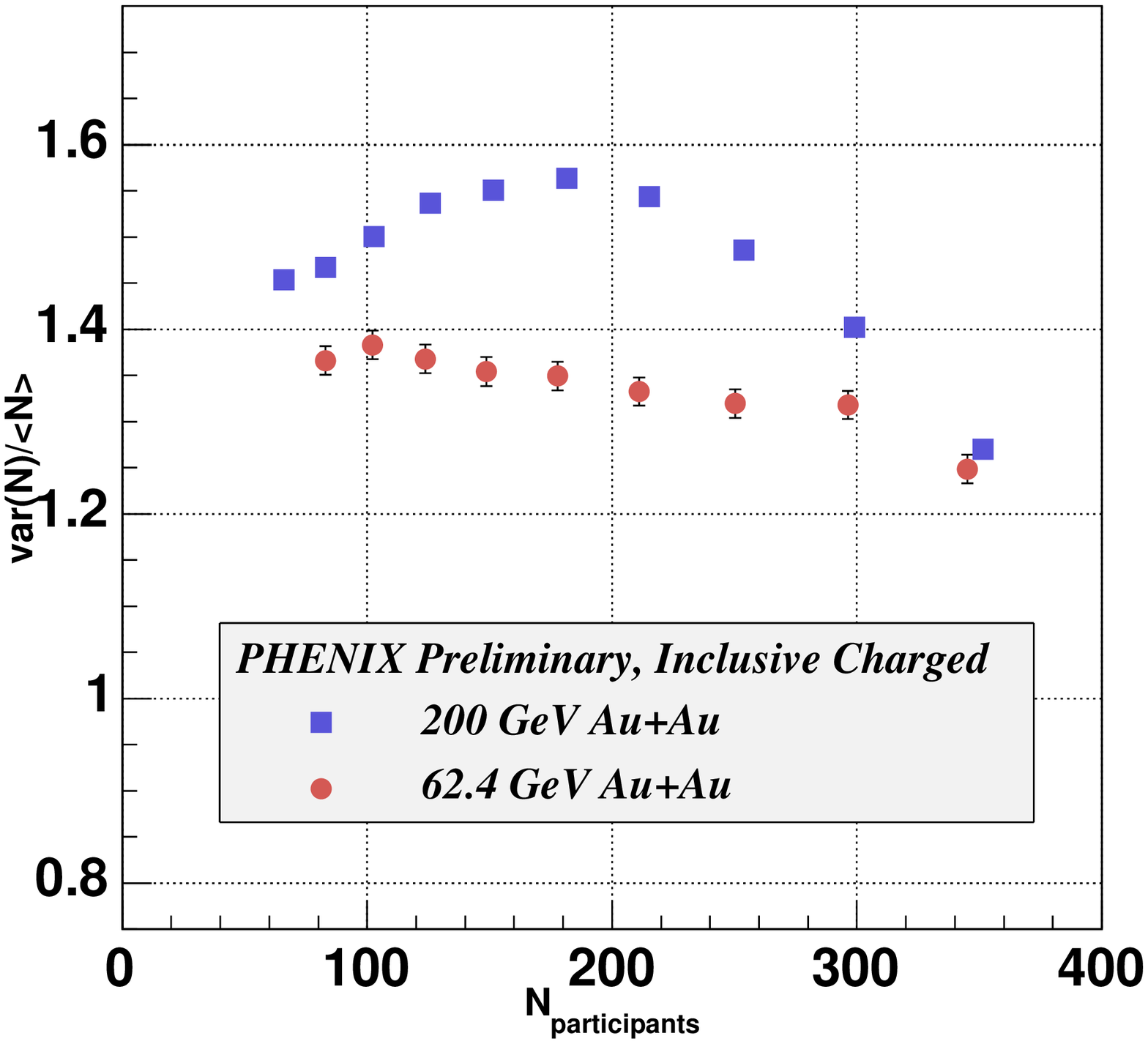}
\caption{\label{fig:mfVsCentAll}Inclusive charged particle multiplicity fluctuations in terms of the scaled variance as a function of centrality for $\sqrt{s_{NN}}$ = 62 and 200 GeV Au+Au collisions. The error bars include statistical and systematic errors.}
\end{minipage}\hspace{2pc}%
\begin{minipage}{18pc}
\includegraphics[width=18pc]{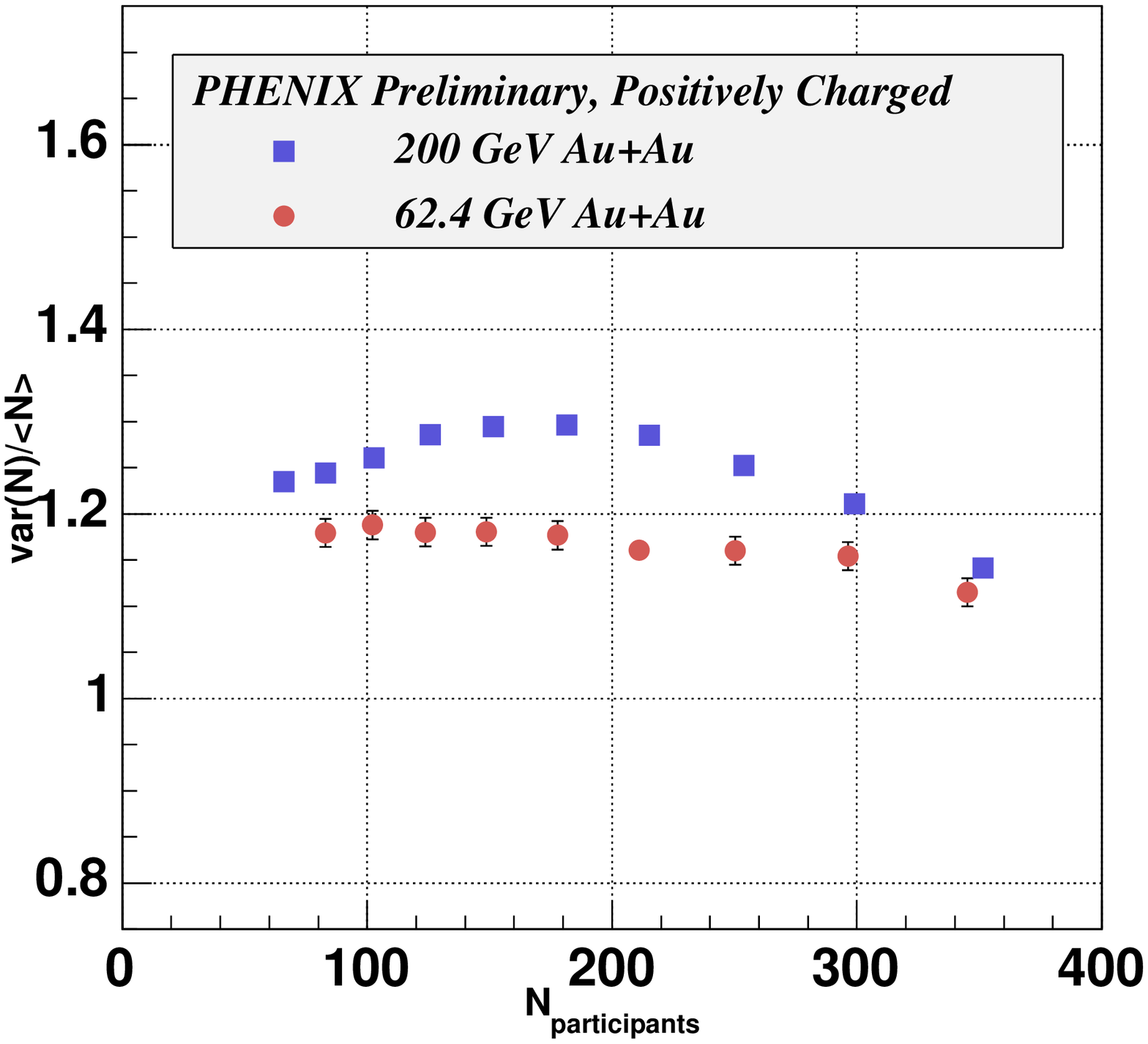}
\caption{\label{fig:mfVsCentPos}Positively charged particle multiplicity fluctuations in terms of the scaled variance as a function of centrality for $\sqrt{s_{NN}}$ = 62 and 200 GeV Au+Au collisions. The error bars include statistical and systematic errors.}
\end{minipage}\hspace{2pc}%
\begin{minipage}{18pc}
\includegraphics[width=18pc]{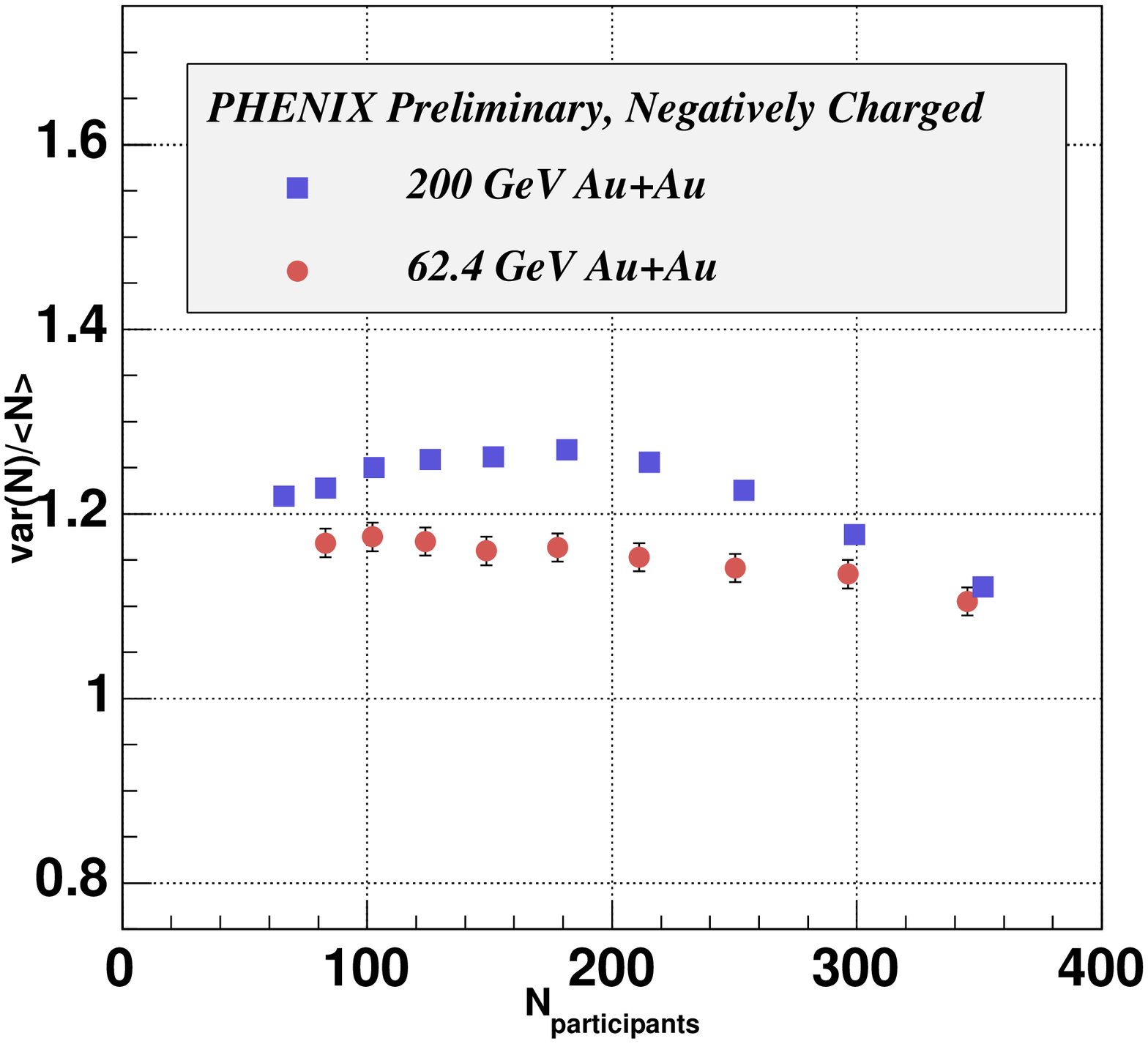}
\caption{\label{fig:mfVsCentNeg}Negatively charged particle multiplicity fluctuations in terms of the scaled variance as a function of centrality for $\sqrt{s_{NN}}$ = 62 and 200 GeV Au+Au collisions. The error bars include statistical and systematic errors.}
\end{minipage} 
\end{figure}

\begin{figure}[h]
\begin{minipage}{18pc}
\includegraphics[width=18pc]{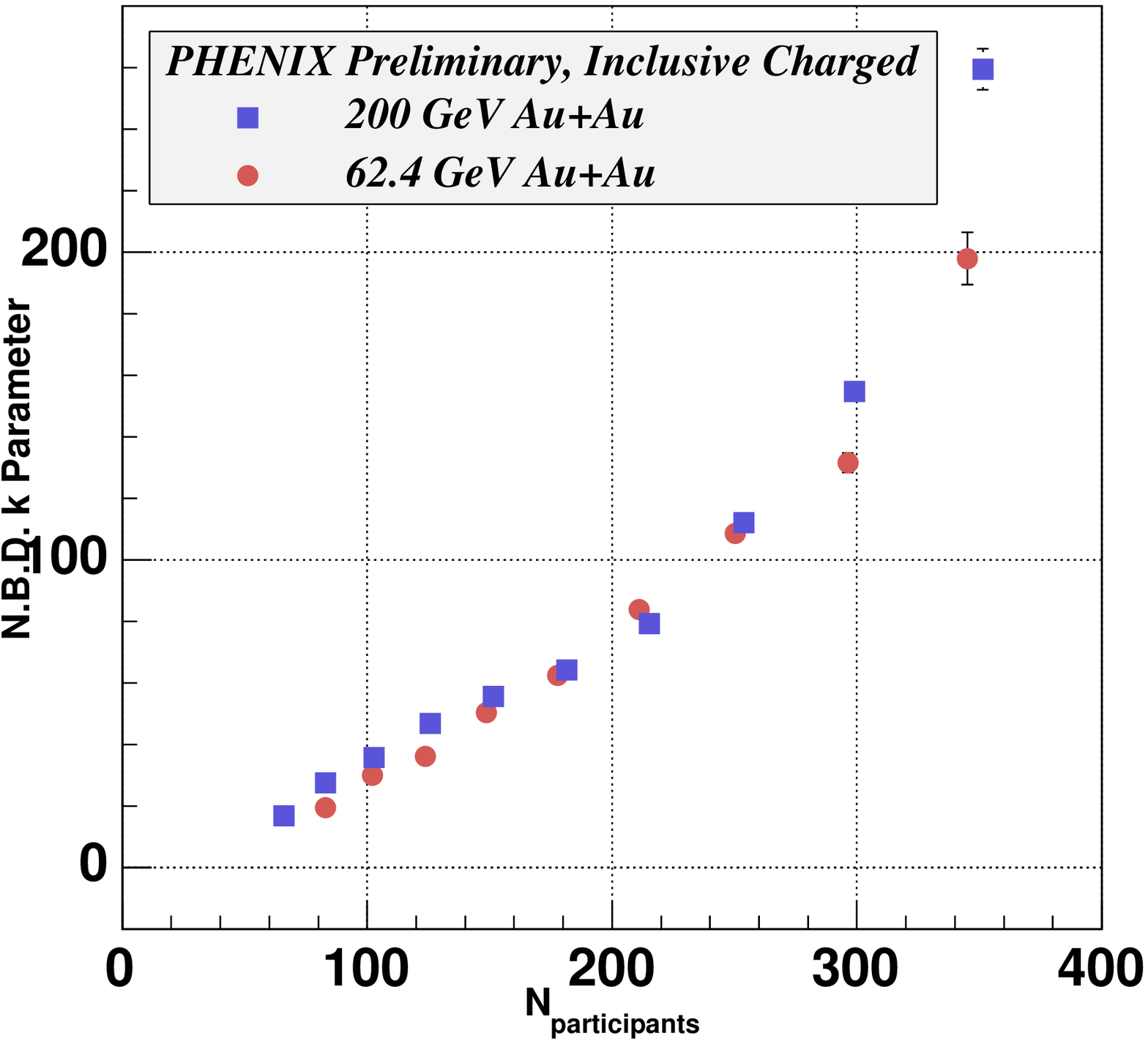}
\caption{\label{fig:mfkVsCentAll}Inclusive charged particle multiplicity fluctuations in terms of the k parameter from a negative binomial distribution fit to the data as a function of centrality for $\sqrt{s_{NN}}$ = 62 and 200 GeV Au+Au collisions.}
\end{minipage}\hspace{2pc}%
\begin{minipage}{18pc}
\includegraphics[width=18pc]{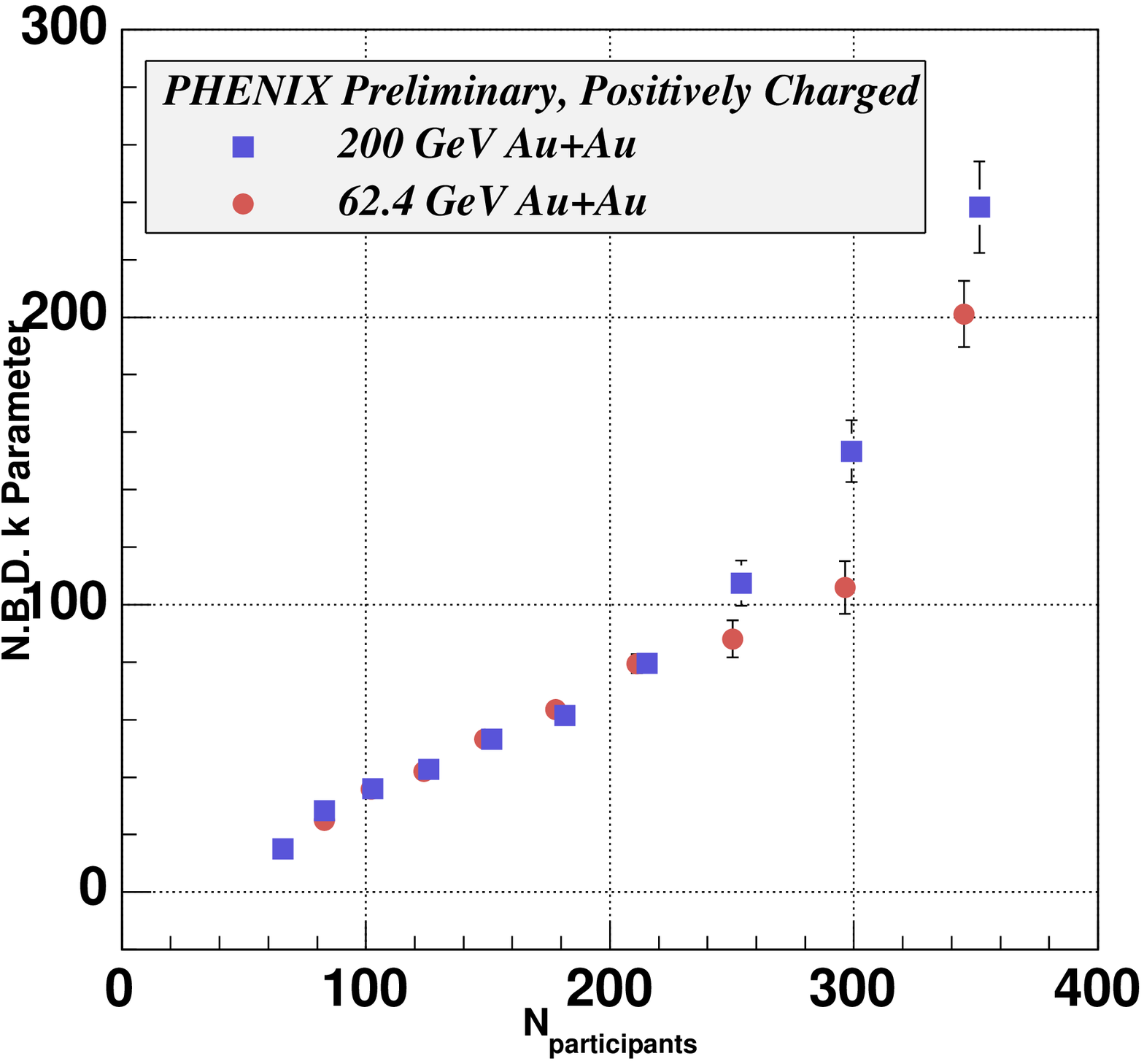}
\caption{\label{fig:mfkVsCentPos}Positively charged particle multiplicity fluctuations in terms of the k parameter from a negative binomial distribution fit to the data as a function of centrality for $\sqrt{s_{NN}}$ = 62 and 200 GeV Au+Au collisions.}
\end{minipage}\hspace{2pc}%
\begin{minipage}{18pc}
\includegraphics[width=18pc]{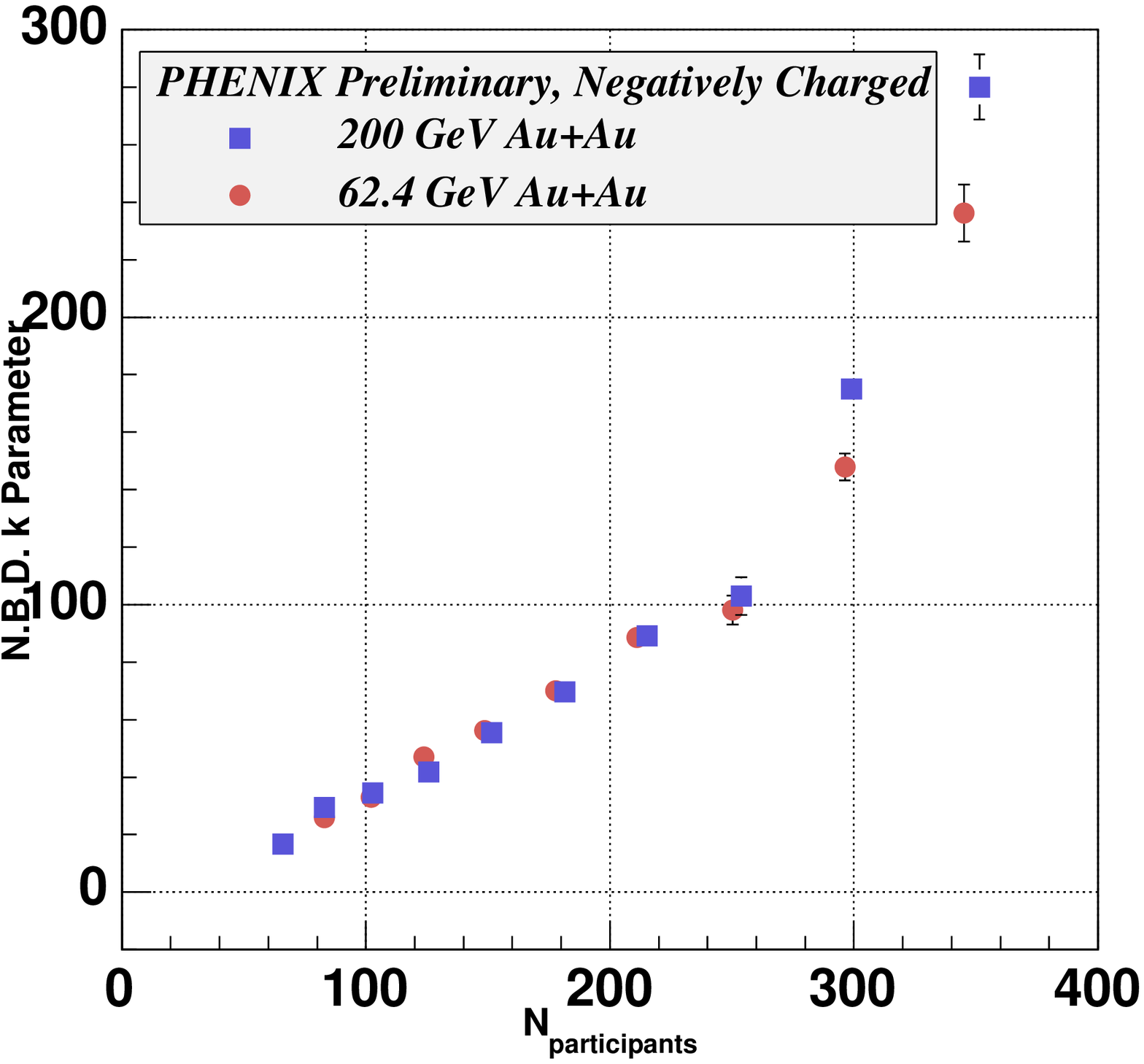}
\caption{\label{fig:mfkVsCentNeg}Negatively charged particle multiplicity fluctuations in terms of the k parameter from a negative binomial distribution fit to the data as a function of centrality for $\sqrt{s_{NN}}$ = 62 and 200 GeV Au+Au collisions.}
\end{minipage} 
\end{figure}

\begin{figure}[h]
\begin{minipage}{18pc}
\includegraphics[width=18pc]{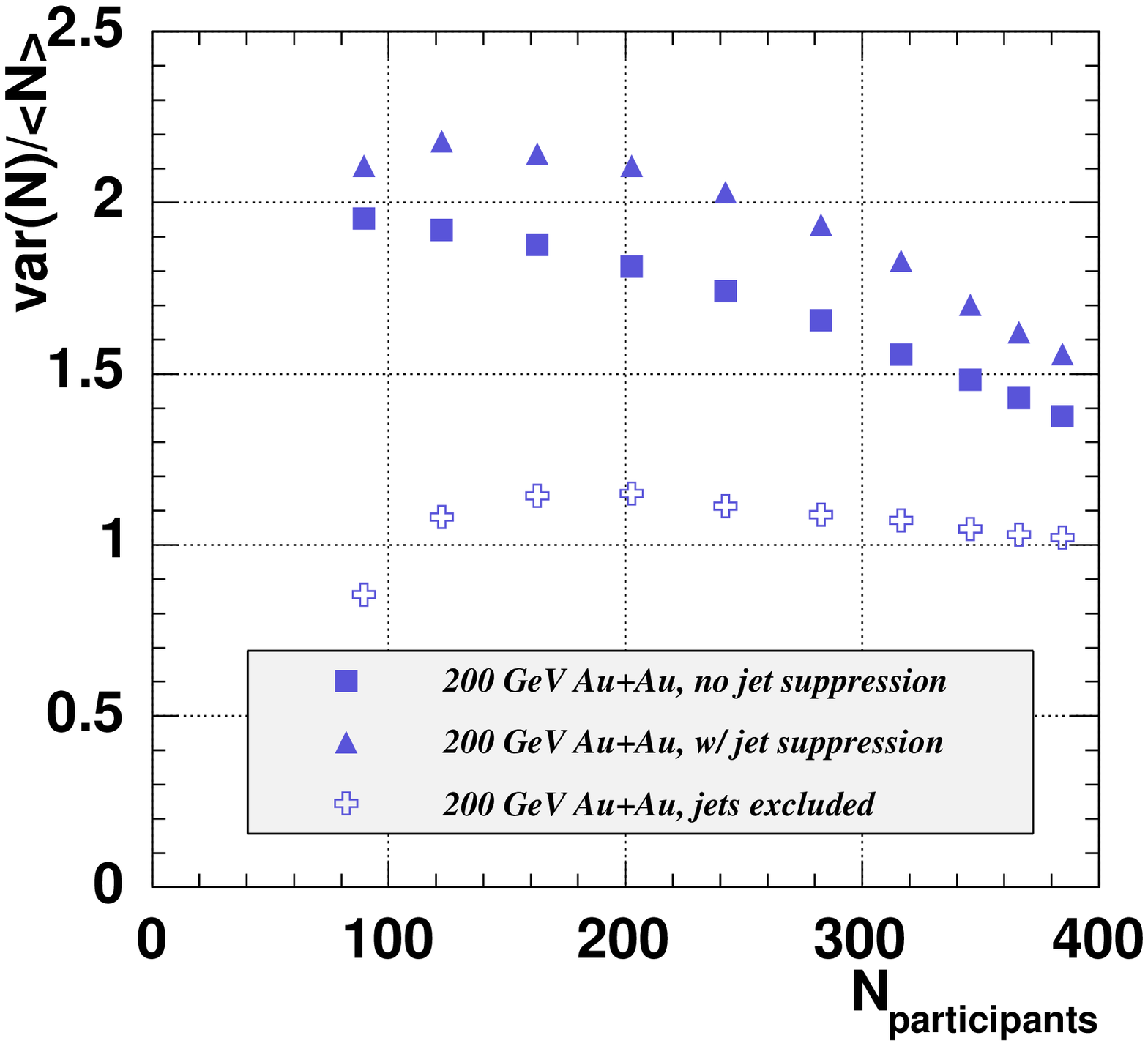}
\caption{\label{fig:mfVsCentHij}Results from a Hijing 1.35 simulation filtered through the PHENIX acceptance for inclusive charged particle multiplicity fluctuations in terms of the scaled variance as a function of $N_{participants}$ for $\sqrt{s_{NN}}$ = 200 GeV Au+Au collisions. Shown are results of the simulation with the default jet energy loss activated, without energy loss, and with jet production deactivated. Note that $<N>$ differs greatly for these 3 cases.}
\end{minipage}\hspace{2pc}%
\begin{minipage}{18pc}
\includegraphics[width=18pc]{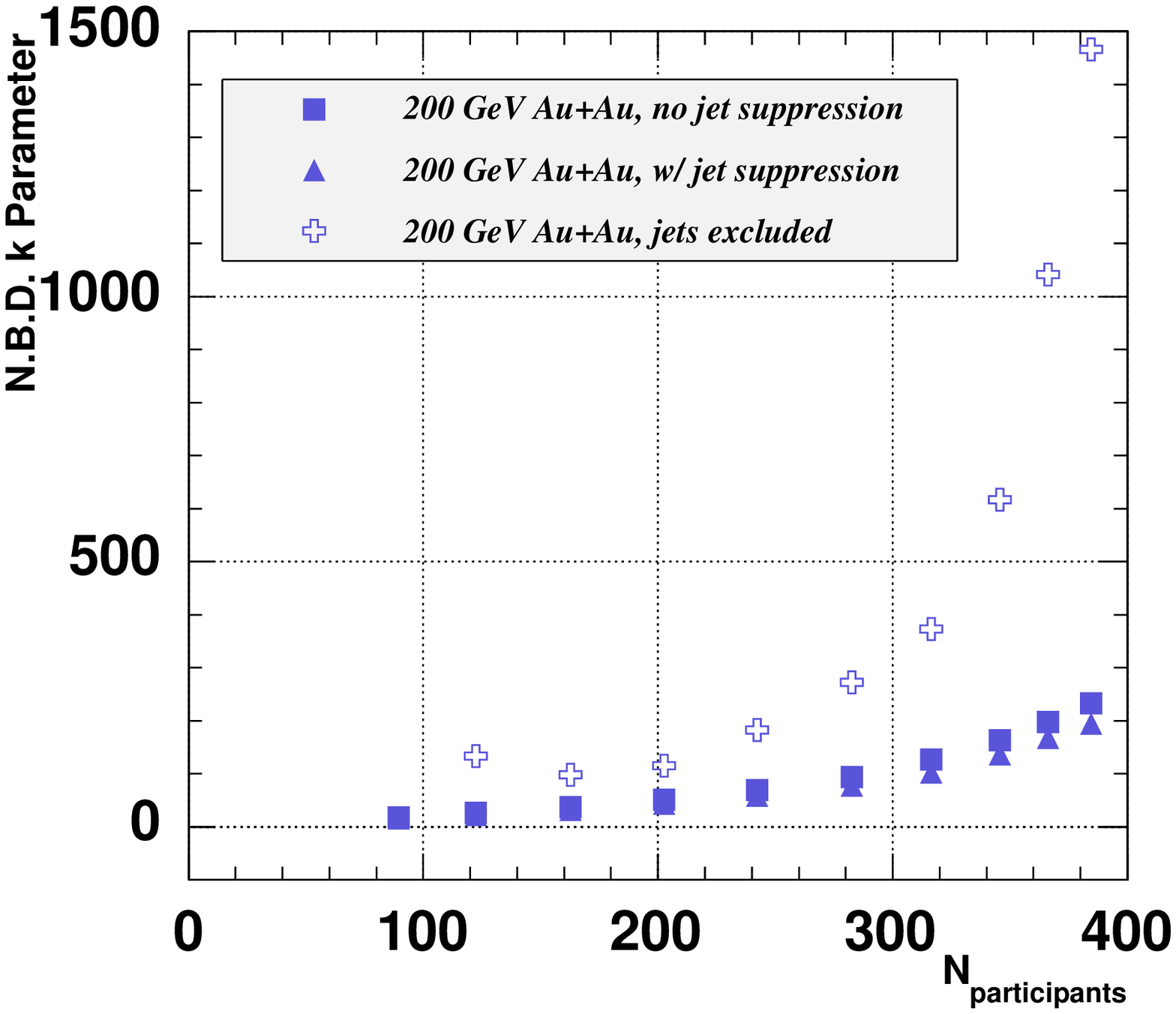}
\caption{\label{fig:mfkVsCentHij}Results from a Hijing 1.35 simulation filtered through the PHENIX acceptance for inclusive charged particle multiplicity fluctuations in terms of the k parameter from a negative binomial distribution fit as a function of $N_{participants}$ for $\sqrt{s_{NN}}$ = 200 GeV Au+Au collisions. Shown are results of the simulation with the default jet energy loss activated, without energy loss, and with jet production deactivated.}
\end{minipage}\hspace{2pc}%
\end{figure}

\begin{figure}[h]
\begin{minipage}{18pc}
\includegraphics[width=18pc]{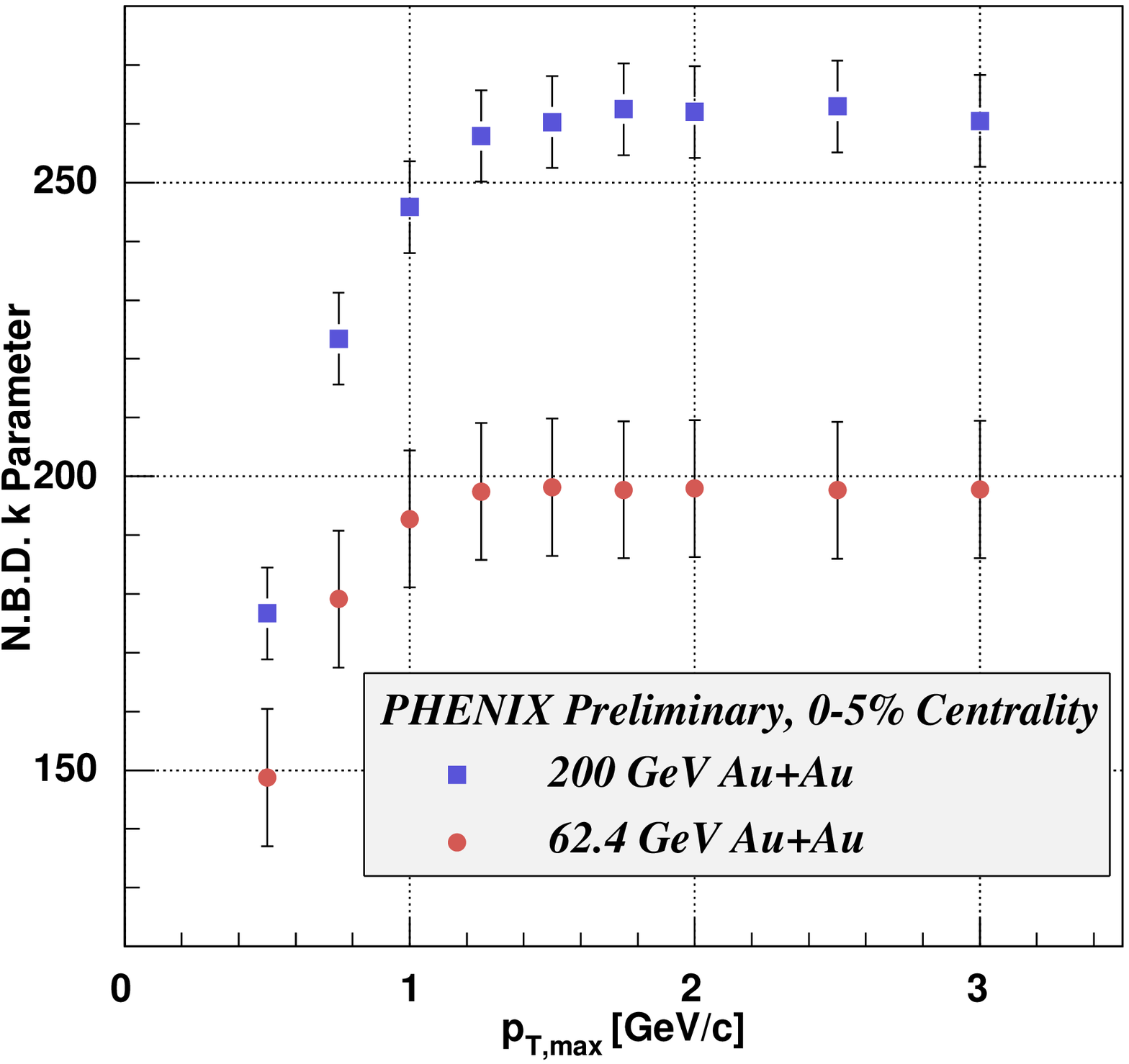}
\caption{\label{fig:mfkVsPtC0}Inclusive charged particle multiplicity fluctuations in terms of the k parameter from a negative binomial distribution fit to the data as a function of $p_{T}$ over the range 200 MeV/c $<p_{T}<p_{T,max}$ for $\sqrt{s_{NN}}$ = 62 and 200 GeV Au+Au 0-5\% central collisions.}
\end{minipage}\hspace{2pc}%
\begin{minipage}{18pc}
\includegraphics[width=18pc]{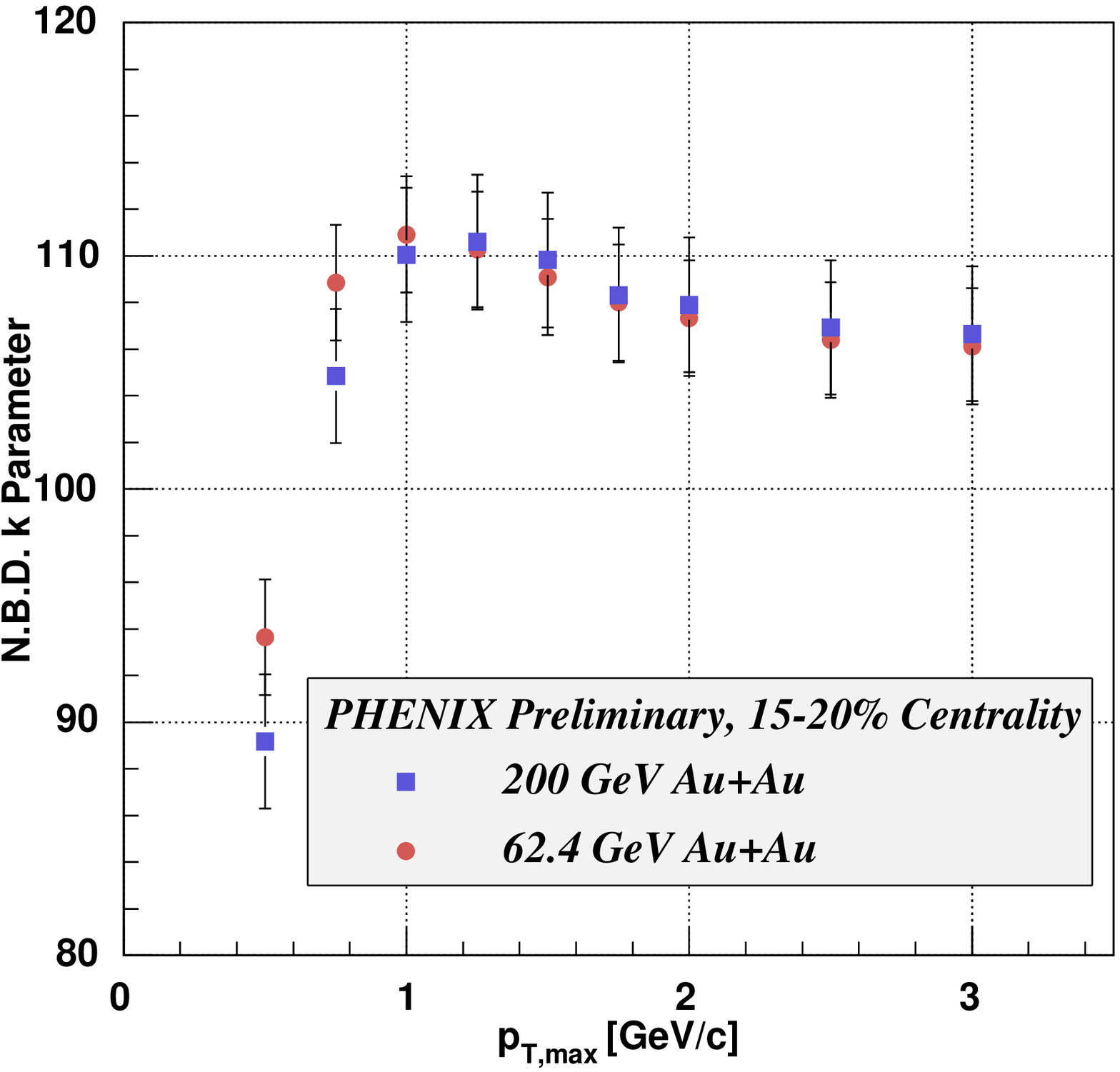}
\caption{\label{fig:mfVskVsPtC2}Inclusive charged particle multiplicity fluctuations in terms of the k parameter from a negative binomial distribution fit to the data as a function of $p_{T}$ over the range 200 MeV/c $<p_{T}<p_{T,max}$ for $\sqrt{s_{NN}}$ = 62 and 200 GeV Au+Au 10-15\% central collisions.}
\end{minipage}\hspace{2pc}%
\begin{minipage}{18pc}
\includegraphics[width=18pc]{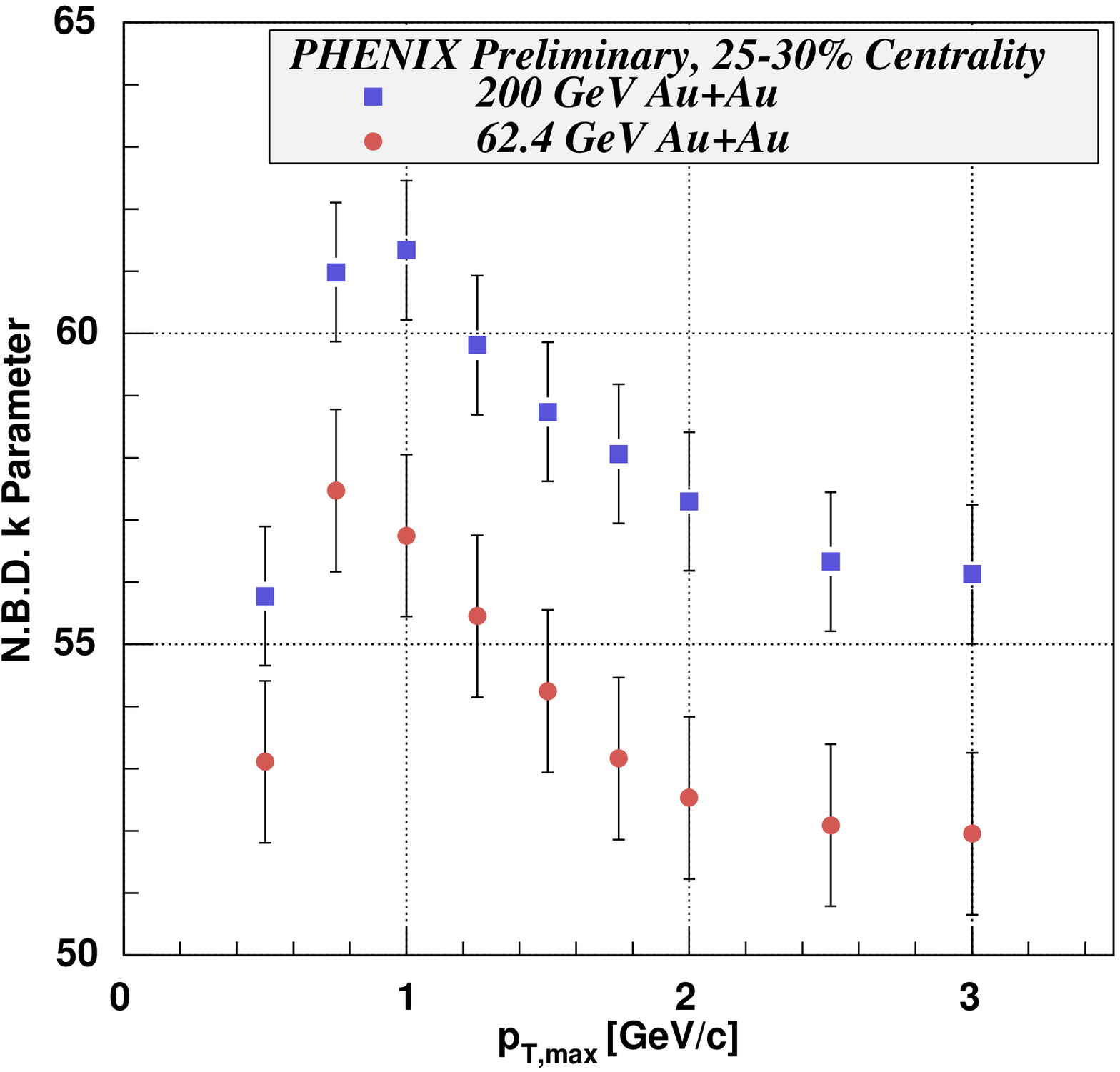}
\caption{\label{fig:mfkVsPtC5}Inclusive charged particle multiplicity fluctuations in terms of the k parameter from a negative binomial distribution fit to the data as a function of $p_{T}$ over the range 200 MeV/c $<p_{T}<p_{T,max}$ for $\sqrt{s_{NN}}$ = 62 and 200 GeV Au+Au 25-30\% central collisions.}
\end{minipage} 
\end{figure}

Charged particle multiplicity fluctuations in terms of the scaled variance as a function of centrality are shown in Fig. \ref{fig:mfVsCentAll} for inclusive charged particles, Fig. \ref{fig:mfVsCentPos} for positive particles, and Fig. \ref{fig:mfVsCentNeg} for negative particles. Contributions from geometry fluctuations are included in these measurements. While the 62 GeV fluctuations are relatively flat as a function of centrality, the 200 GeV data exhibit a structure that peaks in mid-central collisions. The behavior of the fluctuations at both energies are qualitatively different than that observed at the SPS, where the scaled variance increases from central to peripheral collisions over this range. The qualitative features of the data at both of the RHIC energies are the same for inclusive charge, positively charged, and negatively charged particle distributions. Charge selection at RHIC energies appears only to divide the value of $(var(N)/<N>)-1$ roughly by a factor of two.

It is well established that charged particle multiplicity fluctuation distributions in elementary and heavy ion collisions are well described by negative binomial distributions (NBD) \cite{e802MF}. The NBD of an integer $m$ is defined by
\begin{equation}
P(m) = \frac{(m+k-1)!}{m!(k-1)!} \frac{(\mu/k)^{m}}{(1+\mu/k)^{m+k}}
\end{equation}
where $P(m)$ is normalized for $0\leq m \leq \infty$, $\mu\equiv<m>$. The NBD contains an additional parameter, $k$, when compared to a Poisson distribution. The NBD becomes a Poisson distribution in the limit $k\rightarrow\infty$ - the larger the value of $k$, the more Poisson-like is the distribution. The variance and the mean of the NBD is related to $k$ by $1/k = \sigma^{2}/\mu^{2} - 1/\mu$, where $\mu$ is the mean and $\sigma$ is the standard deviation of the distribution. The PHENIX multiplicity distributions are well described by NBD fits for all centralities. The extracted values of the $k$ parameter are shown in Fig. \ref{fig:mfkVsCentAll}-\ref{fig:mfkVsCentNeg} as a function of centrality for inclusive charged, positively charged, and negatively charged particles. Except for the most central collisions, the values of $k$ for the 62 and 200 GeV data are consistent with each other for all charge selections. There is also little difference in $k$ for the different charge selections and the $k$ parameter sharply increases in the most central collisions for both collision energies.

Some of the qualitative features seen in the data can be studied by examining results from a preliminary HIJING simulation within the PHENIX acceptance. Fig. \ref{fig:mfVsCentHij} shows the scaled variance according to HIJING for the following conditions: a) jet energy loss activated, b) jet energy loss deactivated, and c) jet production deactivated. According to this simulation, the activation of jet production, which produces events that are correlated in multiplicity, may explain the rise in the fluctuations seen in the 200 GeV data when going from central to mid-central collisions. However, HIJING over-predicts the magnitude of the scaled variance. Since the jet production cross section is much lower in 62 GeV collisions, their contribution should be greatly reduced, thus the absence of significant jet contributions may explain why the fluctuations are roughly independent of centrality. Shown in Fig. \ref{fig:mfkVsCentHij} are the NBD $k$ parameter results from the HIJING simulation for 200 GeV Au+Au collisions. With jet production excluded, the distributions are well-described by Poisson distributions - the $k$ parameter is very large. Adding jet production introduces a non-Poissonian component to the distribution and reduces $k$ to a value that is close to that of the data. There is little difference when including or excluding energy loss.

In order to better understand the various sources of the multiplicity fluctuations, the fluctuations have been measured as a function of $p_T$ range, $200~MeV/c<p_{T}<p_{T,max}$. The results in terms of the NBD $k$ parameter are shown for three centralities in Fig. \ref{fig:mfkVsPtC0}-\ref{fig:mfkVsPtC5}. The qualitative behavior of the fluctuations changes dramatically when comparing central to mid-central collisions. This behavior may be consistent with what is expected from jet suppression. In the most central collisions, jet suppression effectively dilutes the hard, non-Poissonian contribution to the distribution resulting in an increasing value of $k$. In non-central collisions, the dilution of the non-Poissonian hard component is reduced and $k$ decreases. This could also explain the rapid increase in $k$ as a function of centrality in the most central collisions. More detailed simulations are being studied to better understand the measured trends and to better estimate contributions from hard processes and geometry fluctuations.

\section{Summary}

PHENIX has measured fluctuations of charge, mean transverse momentum, and charged particle multiplicity. PHENIX has demonstrated that despite the small acceptance, sensitive and accurate measurements of fluctuation observables can still be made. The charge fluctuation measurements are consistent with statistically independent particle emission. The mean transverse momentum fluctuation measurements demonstrate that any residual temperature fluctuations that may be present are very small, and are on the order of the fluctuations seen at SPS energies. The multiplicity fluctuation measurements demonstrate a qualitative behavior as a function of centrality that differs from measurements at SPS energies. There are interesting prospects for future fluctuation measurements from PHENIX using the new Cu+Cu datasets taken at $\sqrt{s_{NN}}$=19, 62, and 200 GeV.

\smallskip


\section{References}
\medskip
\begin{thebibliography}{9}
\bibitem{phenixNIM} Adcox K {\it et al} (PHENIX Collaboration) 2003 {\it Nucl. Instrum. Meth.} A {\bf 499} 469.
\bibitem{ppg005} Adcox K {\it et al} (STAR Collaboration) 2002 {\it Phys. Rev.} C {\bf 66} 024901.
\bibitem{Jeo00} Jeon S and Koch V 2000 {\it Phys. Rev. Lett.} {\bf 85} 2076.
\bibitem{phxQfluc130} Adcox K {\it et al} (PHENIX Collaboration) 2002 {\it Phys. Rev. Lett.} {\bf 89} 082301.
\bibitem{phxQfluc200} Nystrand J {\it et al} (PHENIX Collaboration) 2003 {\it Nucl. Phys.} A {\bf 715} 603c.
\bibitem{Bia02} Bialas A 2002 {\it Phys. Lett.} B {\bf 532} 249.
\bibitem{Ste99} Stephanov M {\it et al} 1999 {\it Phys. Rev.} D {\bf 60} 114028.
\bibitem{Gav04} Gavin S 2004 {\it Phys. Rev. Lett.} {\bf 94} 162301.
\bibitem{ppg027} Adcox K {\it et al} (PHENIX Collaboration) 2004 {\it Phys. Rev. Lett.} {\bf 93} 092301.
\bibitem{starPtFluc} Adams J {\it et al} (STAR Collaboration) 2003 {\it Preprint} nucl-ex/0308033.
\bibitem{jtmQM2004} Mitchell J 2004 {\it J. Phys.} G {\bf 30} S819.
\bibitem{Kor01} Korus R 2001 {/it Phys. Rev.} C {\bf 64} 054908.
\bibitem{na49MF} Rybczynski M {\it et al} (NA49 Collaboration) 2004 {\it Preprint} nucl-ex/0409009.
\bibitem{e802MF} Abbott T {\it et al} (E-802 Collaboration) 1995 {\it Phys. Rev.} C {\bf 52} 2663.
\end{thebibliography}
\end{document}